\documentclass[printer]{aa}

\date{June 26, 2026}

\usepackage{amsmath}

\usepackage{graphicx}

\usepackage{txfonts}

\usepackage[colorlinks,citecolor=blue]{hyperref}
\usepackage{orcidlink}
\usepackage{threeparttable}

\graphicspath{{./}}

\begin{document}

 \title{Euclid Q1 reveals spatial variations of the extinction law in the dense cloud LDN 1641}

   \author{Rui Chen\orcidlink{0009-0002-7197-3267}\inst{1}
        \and Shu  Wang\orcidlink{0000-0003-4489-9794}\inst{2}\fnmsep\inst{3}\fnmsep\thanks{Corresponding author: shuwang@nao.cas.cn}
        \and Xiaodian Chen\orcidlink{0000-0001-7084-0484}\inst{2}\fnmsep\inst{3}\fnmsep\inst{4}
        \and Kun Wang\orcidlink{0000-0002-5745-827X}\inst{1}   }

   \institute{School of Physics and Astronomy, China West Normal University, Nanchong 637009, China
   \and CAS Key Laboratory of Optical Astronomy, National Astronomical Observatories, Chinese Academy of Sciences, Beijing 100101, China
   \and School of Astronomy and Space Science, University of the Chinese Academy of Sciences, Beijing 101408, China
   \and Institute for Frontiers in Astronomy and Astrophysics, Beijing Normal University, Beijing 102206, China}

  \abstract 
   {Dust extinction laws are essential for precision photometry and provide a direct probe of grain properties, but their behavior in dense molecular clouds remains poorly constrained at high extinction. Using Euclid Quick Data Release 1 (Q1) imaging of the Orion~A dark cloud Lynds Dark Nebula 1641 (LDN~1641), we measure the extinction law from the broad Visible Instrument (VIS) band and the Near-Infrared Spectrometer and Photometer (NISP) $Y$, $J$, and $H$ bands along sightlines reaching $A_V\sim 30$~mag toward the cloud core. We derive color-excess ratios $E(\lambda-H)/E(Y-H)$ from linear fits to color--color diagrams of $(\lambda-H)$ versus $(Y-H)$, and convert them into relative extinctions $A_\lambda/A_H$. The near-infrared extinction in LDN~1641 is well described by a power law, $A_\lambda\propto \lambda^{-\alpha}$, with $\alpha=1.57 \pm 0.06$, corresponding to $A_{\rm VIS}/A_H=4.23 \pm 0.24$, $A_Y/A_H=2.13 \pm 0.18$, and $A_J/A_H=1.47 \pm 0.11$. We further find significant spatial variations: $\alpha$ changes by up to $27\%$, with systematically smaller $\alpha$ values, and therefore flatter extinction curves, in higher-extinction regions. This flattening is consistent with an enhanced large-grain population and supports substantial grain growth from the diffuse outskirts to the dense core of a single molecular cloud.}


   \titlerunning{Euclid Q1 reveals spatial variations of the extinction law in the dense cloud LDN 1641}
   \authorrunning{Chen et al.}

   \maketitle
   \nolinenumbers

\section{Introduction}
\label{sec:intro}

The extinction law affects the accuracy of reddening corrections and provides direct constraints on dust evolution. Its spatial variation within dense molecular clouds is therefore a key observational test of grain-growth models \citep{2003ARA&A..41..241D}. For quantitative comparison, the extinction law is commonly described using color-excess ratios (CERs), such as $E(\lambda-V)/E(B-V)$, and relative extinctions, such as $A_\lambda/A_V$, where $E(\lambda-V)$ is the color excess and $A_\lambda$ is the absolute extinction at wavelength $\lambda$ \citep{1989ApJ...345..245C,Wang2019}. In the ultraviolet (UV) and optical, extinction curves have been measured extensively and are often parameterized by the total-to-selective extinction ratio $R_V = A_V/E(B-V)$ \citep{1989ApJ...345..245C,1999PASP..111...63F,Wang2019,2020ApJ...891...67M}. In the near-infrared (NIR), the extinction law is usually approximated by a power law, $A_\lambda \propto \lambda^{-\alpha}$ \citep{1990ARA&A..28...37M, 2003ARA&A..41..241D, 2024ApJ...964L...3W}. Whether this form is universal, and whether the power-law index $\alpha$ varies systematically with environment, remains debated. Several studies find that a power law provides a robust description of NIR extinction along many sightlines \citep{2005ApJ...619..931I,2014ApJ...788L..12W,2019A&A...630L...3N}, whereas others report substantial differences in $\alpha$ between environments \citep{2006ApJ...638..839N, 2009MNRAS.394.2247G}. Dense molecular-cloud cores are especially important because coagulation and ice-mantle accretion can modify the effective grain-size distribution and optical properties. These processes are supported by theoretical models \citep{1994A&A...291..943O,2011A&A...532A..43O} and observational evidence \citep{2015A&A...584A..93J, li2024flattestinfraredextinctioncurve}; for example, \citet{li2024flattestinfraredextinctioncurve} found that the infrared extinction curve may flatten in dense regions, corresponding to a decrease in $\alpha$. Overall, reported values of $\alpha$ span a wide range, from about $1.6$ to $2.6$, across different sightlines and environments \citep{1985ApJ...288..618R,1989ApJ...345..245C,2005A&A...435..575M,2009MNRAS.400..731S,2011ApJ...737...73F,Wang2019,2024ApJ...964L...3W,2017ApJ...849L..13A,2020MNRAS.496.4951M,2010A&A...511A..18S}. Recently, \citet{2024ApJ...964L...3W} used spectra from the James Webb Space Telescope (JWST) Near-Infrared Spectrograph (NIRSpec) to derive an average infrared extinction power-law index for the Milky Way of $\alpha = 1.98 \pm 0.15$, and suggested that spatial variations may be related to factors such as regional density structure and star-formation activity. Systematic measurements of the spatial variation of $\alpha$ are therefore essential for understanding dust evolution.

Orion~A is one of the most active nearby star-forming complexes in the Milky Way. Its multi-scale structure, from diffuse outskirts to dense cores, provides an ideal laboratory for studying extinction laws and dust evolution. \citet{2021ApJ...915...74U} combined optical and Two Micron All Sky Survey (2MASS) NIR photometry and found that $R_V$ around Orion~A is significantly higher than the typical value in the diffuse interstellar medium, placing constraints on the upper limit of grain size. \citet{2024AJ....168..256C} combined stellar parameters from the Large Sky Area Multi-Object Fiber Spectroscopic Telescope (LAMOST) and the Apache Point Observatory Galactic Evolution Experiment (APOGEE) with multi-band photometry and showed that infrared extinction in dense molecular-cloud regions may deviate from the average behavior because of effects such as ice absorption. However, accurately tracing the continuous spatial evolution of extinction laws from diffuse environments to dense cores within a single cloud still requires high spatial resolution and deep multi-band photometry.

The Quick Data Release 1 (Q1) from the Euclid mission provides a new opportunity to address these questions. The Euclid Visible Instrument (VIS) operates over a broad bandpass ($550$ -- $900$~nm) with an exceptional spatial resolution of $\sim 0.1^{\prime\prime}$ per pixel. The Near-Infrared Spectrometer and Photometer (NISP) provides photometry in three broad bands ($Y$, $J$, $H$) with a pixel scale of 0.3$^{\prime\prime}$, reaching a 5$\sigma$ point-source limiting magnitude of $H_{\rm AB}\simeq 24.4$ mag \citep{euclidcollaboration2025euclidquickdatarelease}. This depth enables the detection of many faint sources along high-extinction sightlines. The Q1 dataset includes observations of Lynds Dark Nebula 1641 (LDN~1641; also referred to as L1641) within the Orion A complex, providing a valuable sample for investigating extinction laws in dense molecular clouds \citep{2025arXiv250315302E}. LDN~1641 exhibits pronounced extinction gradients, from $A_V \sim 1$ -- $3$~mag at its periphery to $\gtrsim 20$~mag in its central regions. It is characterized by an intricate network of filaments and dense cores, and hosts a rich population of young stellar objects, protostars, Herbig-Haro objects, and molecular outflows \citep{2011A&A...527A..60L,2018A&A...614A..65M}. LDN~1641 is therefore an ideal laboratory for probing extinction laws and dust-grain evolution.

In this paper, we systematically derive the optical-to-NIR extinction law in LDN~1641 using high-precision Euclid Q1 VIS and NISP ($Y$, $J$, $H$) photometry. We also divide the cloud into subregions by extinction level to investigate spatial variations. The paper is organized as follows. Section~\ref{sec:Data} describes the data and sample selection. Section~\ref{sec:method} presents the method and core measurements of CERs and relative extinction. Section~\ref{sec:discussion} discusses the environmental dependence of the extinction law. Section~\ref{sec:summary} summarizes the main conclusions.

\section{Data and sample}
\label{sec:Data}

Euclid Q1 covers three deep fields, Euclid Deep Field North (EDF-N), Euclid Deep Field South (EDF-S), and Euclid Deep Field Fornax (EDF-F), as well as the LDN~1641 nebula. In this work, we focus on LDN~1641.
We use the broad Euclid VIS band together with the NISP $Y$, $J$, and $H$ bands, with all magnitudes on the AB magnitude system.

\begin{figure*}[htbp]
    \centering
    \includegraphics[width=1\linewidth]{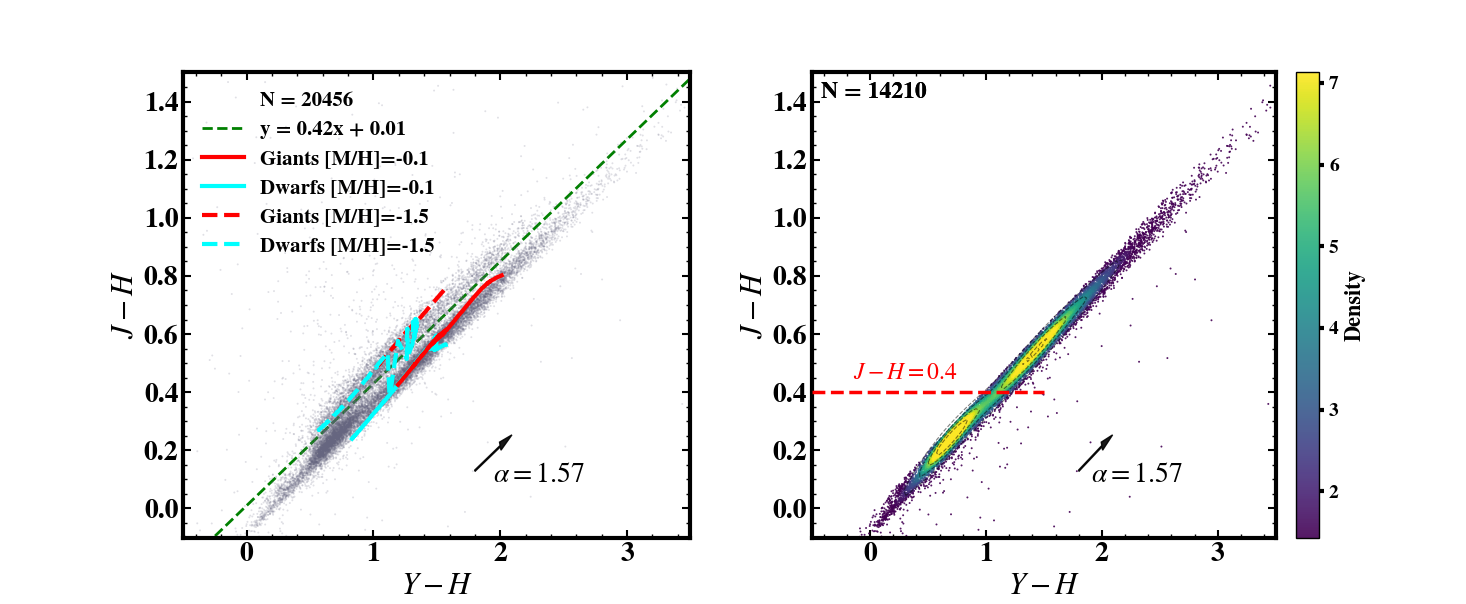}
   \caption{Euclid $(J-H)$ versus $(Y-H)$ color--color diagrams for sources in the LDN~1641 field. Left: Scatter plot for $N=20456$ sources after photometric-quality cuts, showing the bifurcated source distribution. For guidance, we overplot illustrative giant and dwarf evolutionary tracks for [M/H]$=-0.1$ and $-1.5$, reddened assuming a power-law extinction curve with $\alpha=1.57$ and $A_V=7$ mag. The black arrow shows an illustrative reddening vector for $\alpha=1.57$. The green dashed line, defined by $(J-H)=0.42\,(Y-H)+0.01$, follows the low-density valley between the two branches. We retain the sources below this line, corresponding to the lower reddened sequence with smaller $(J-H)$ at a given $(Y-H)$; this sequence is more populated and is more consistent with the relatively metal-rich population expected for the Orion~A star-forming complex. Right: Density distribution for the $N=14210$ sources retained after this branch selection. The remaining sources show a low-$(J-H)$ concentration associated with low-extinction and/or foreground objects and a more highly reddened sequence extending to larger $(Y-H)$ and $(J-H)$. The red horizontal dashed line at $J-H=0.4$ mag marks the additional threshold used to remove the low-extinction/foreground concentration.}
    \label{fig:ccd}
\end{figure*}

To ensure reliable color measurements, we first require magnitude uncertainties of $\sigma_m < 0.05$ mag in all bands. Sample selection is then performed in the $(J-H)$--$(Y-H)$ diagram (Figure~\ref{fig:ccd}). The left panel shows a clear bifurcation, indicating stellar populations with different intrinsic colors. To illustrate its likely origin, we overplot giant and dwarf evolutionary tracks for [M/H]$=-0.1$ and $-1.5$, reddened with a power-law extinction curve of $\alpha=1.57$ and $A_V=7$ mag. These tracks qualitatively reproduce the two branches, suggesting that the bifurcation is driven by a combination of metallicity and luminosity-class differences. Since LDN~1641 belongs to the nearby Orion~A star-forming complex, the dominant reddened population is expected to be relatively metal-rich rather than metal-poor. The lower branch is more populated and continuous, and is more consistent with the relatively metal-rich population expected for Orion~A. We therefore adopt this lower branch as the main sample for the extinction-law analysis. The two branches are separated by a clear low-density valley in the two-dimensional source-density distribution. We use the density map and its contours to trace this valley and approximate it with the empirical linear boundary $(J-H)<0.42\,(Y-H)+0.01$ (green dashed line in Figure~\ref{fig:ccd}). This criterion retains the lower reddened sequence, which has smaller $(J-H)$ at a given $(Y-H)$, and reduces contamination from the distinct upper branch; the number of sources is reduced from 20,456 after the photometric-quality cuts to 14,210 after applying this branch selection.
The right panel shows the density distribution after this branch selection. Two concentrations remain: a lower-left low-extinction/foreground component and an upper-right highly reddened component tracing sightlines through LDN~1641. To suppress the former component further and retain the main highly reddened sample, we exclude sources with $J-H<0.4$ mag. This threshold is shown by the red horizontal dashed line in the right panel of Figure~\ref{fig:ccd}.

\begin{figure*}[htbp]
    \centering
    \includegraphics[width=1.0\linewidth]{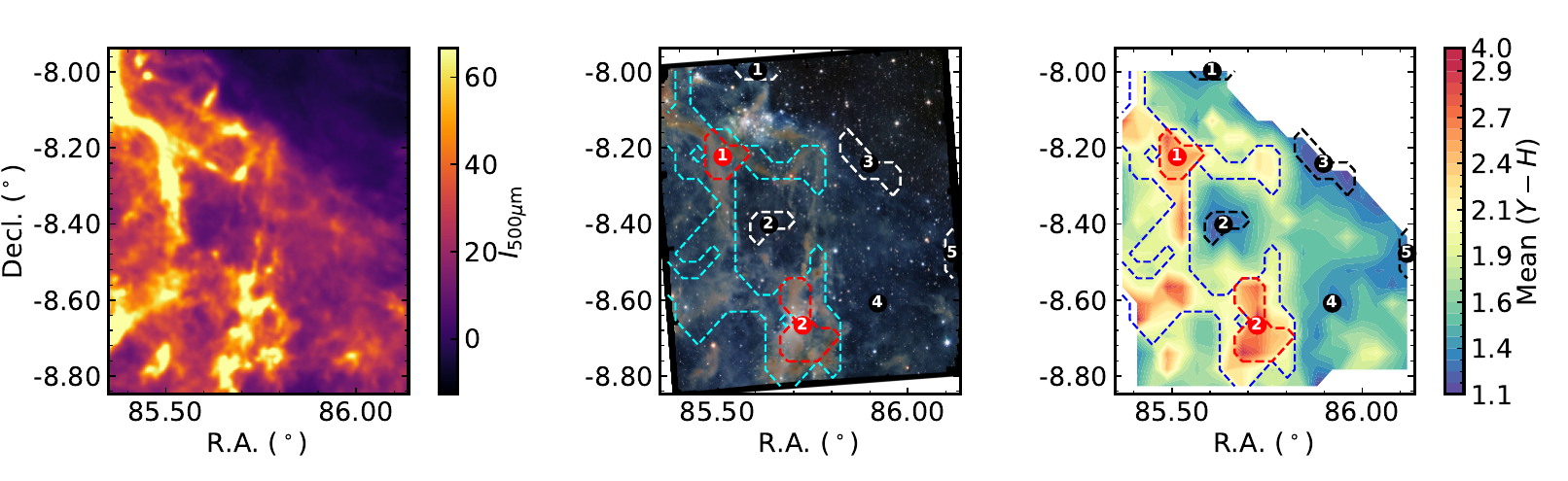}
  \caption{Spatial distributions of Herschel dust emission and the Euclid NIR binned mean color $(Y-H)$ in LDN~1641. 
  Left: Herschel/SPIRE 500~$\mu$m surface brightness map ($I_{500}$; MJy\,sr$^{-1}$), used as an empirical proxy sensitive to dust column density and temperature. 
  Middle: Euclid pseudo-color composite image of the region. 
  Right: Interpolated map of the $20\times20$ binned mean $(Y-H)$ color, serving as a reddening/extinction proxy across the same field. In the middle and right panels, the cyan/blue dashed boundary outlines the final high-extinction study footprint selected from connected bins with mean $(Y-H) > 1.9$ mag and more than 300 sources. The labeled polygons mark subregions used for environmental comparison in Section~\ref{sec:discussion}: high-extinction core subregions (red; IDs~1--2) and low-extinction reference subregions (white/black; IDs~1--5).} 
    \label{fig:herschel}
\end{figure*}

In addition, to ensure that the selected sample primarily traces the high-extinction environment of LDN~1641, we compare the Herschel Spectral and Photometric Imaging Receiver (SPIRE) 500~$\mu$m dust emission map with the Euclid imaging and the spatial distribution of NIR colors (Figure~\ref{fig:herschel}). The left panel of Figure~\ref{fig:herschel} shows the 500~$\mu$m surface brightness $I_{500}$ (MJy\,sr$^{-1}$), which serves as an empirical proxy sensitive to dust column density and temperature. The middle panel presents the Euclid pseudo-color image of the same field, providing a direct view of the cloud morphology in the optical--NIR bands. By comparing these two panels, we find that the upper-right part of the field, approximately within right ascension (R.A.) $85.6^\circ<{\rm R.A.}<86.1^\circ$ and declination (Decl.) $-8.4^\circ<{\rm Decl.}<-8.0^\circ$, is relatively faint in the Herschel map and also lacks the prominent obscuration structures seen elsewhere. In this region, both $I_{500}$ and $(Y-H)$ remain generally low, typically with $I_{500}\lesssim10$ MJy\,sr$^{-1}$ and $(Y-H)\lesssim0.8$ mag, suggesting that this corner is dominated by diffuse outer material or low-extinction sightlines. To avoid contamination from this low-extinction edge while retaining the main high-extinction cloud area, we exclude this corner using an empirical boundary defined as ${\rm R.A.} < -1.157\,({\rm Decl.}+7.965^\circ)+85.594^\circ$. 

After excluding the low-extinction edge region in the upper-right corner, we use the spatial distribution of the $(Y-H)$ color to define the final study region (Figure~\ref{fig:herschel}, right panel). The selected Euclid sources are divided into $20\times20$ spatial bins in the R.A.--Decl. plane, and the mean $(Y-H)$ color is computed for each bin. Bins with mean $(Y-H)>1.9$ mag are selected as high-extinction candidates; adjacent selected bins are then grouped into connected regions, and only regions containing more than 300 sources are retained. The adopted boundary is therefore determined from the original binned mean-$(Y-H)$ values. The blue dashed boundary in the middle and right panels of Figure~\ref{fig:herschel} outlines the final retained study region.
In Section~\ref{sec:discussion}, we further analyze selected subregions to examine the extinction law under different environmental conditions. The red polygons denote two high-extinction core subregions (labeled 1 and 2), while the black polygons indicate five low-extinction reference subregions (labeled 1--5). 

\section{Method and results} 
\label{sec:method} 

We determine the extinction law in three steps. First, we derive the CERs $E(J-H)/E(Y-H)$ and $E({\rm VIS}-H)/E(Y-H)$ from color--color diagrams. Second, we convert the measured $E(J-H)/E(Y-H)$ into the power-law index $\alpha$, assuming that the NIR extinction follows $A_\lambda \propto \lambda^{-\alpha}$. Finally, we use this value of $\alpha$ to compute $A_Y/A_H$ and $A_J/A_H$, and then combine $A_Y/A_H$ with $E({\rm VIS}-H)/E(Y-H)$ to derive $A_{\rm VIS}/A_H$. This procedure yields the relative extinction law of LDN~1641 from optical to NIR wavelengths.
Hereafter, `full sample' refers to the Euclid-selected sources within the final high-extinction study footprint defined in Section~\ref{sec:Data}.

Because the Euclid high-extinction sample lacks star-by-star intrinsic-color information and sufficiently broad multi-band photometric coverage, we do not directly fit total-to-selective extinction ratios as in \citet{1990ApJ...357..113M} and \citet{2005ApJ...623..897L}. Instead, we use the slopes of the color--color diagrams to measure CERs and then convert the NIR ratio into the power-law index $\alpha$.

\subsection{CERs} \label{subsec:CER}

\begin{figure*}[hbtp]
    \centering
    \includegraphics[width=1\linewidth]{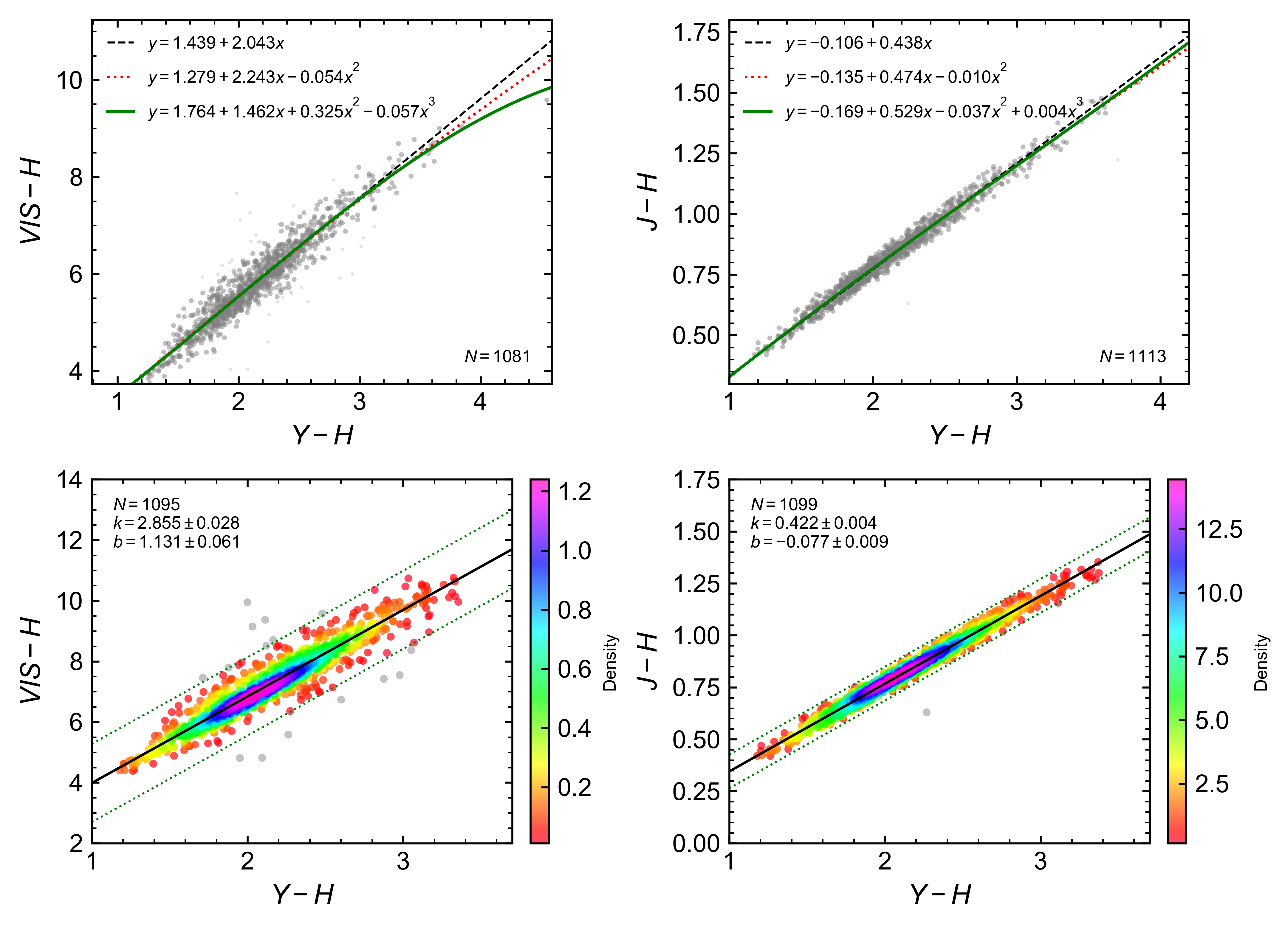}
    \caption{Optical and NIR color--color diagrams and extinction-law fits for the LDN~1641 region. Upper panels: Fits to the observed data without curvature correction. The left and right panels show $(\mathrm{VIS}-H)$ versus $(Y-H)$ and $(J-H)$ versus $(Y-H)$, respectively. The black dashed, red dotted, and green solid lines represent the best-fitting linear, second-order, and third-order polynomials, respectively, illustrating the characteristic curvature. Lower panels: Color--color diagrams after curvature correction. The color shading indicates the source density. Black solid lines show the best linear fits, with the fitting parameters (slope $k$, intercept $b$) and the effective number of sources $N$ annotated in the upper-left corners. Gray points represent 3$\sigma$ outliers excluded from the fitting.}
    \label{fig:curvature}
\end{figure*}

We first constructed the $(J-H)$ versus $(Y-H)$ and $(\mathrm{VIS}-H)$ versus $(Y-H)$ color--color diagrams for the full sample in the final study region. In the NIR bands, the intrinsic color dispersion of red giants is small ($<0.1$~mag; \citealt{1983A&A...128...84K,2011A&A...527A..60L}), so the observed color spread in the selected sample is expected to be dominated mainly by reddening rather than by stellar-type differences. The CER is inferred from the slope of the color--color distribution. Intrinsic-color scatter primarily increases the dispersion around the relation and has a limited effect on the fitted slope, especially given the long reddening baseline provided by the Euclid data. In the optical bands, red giants span a broader range of intrinsic colors ($0.4<(\mathrm{VIS}-H)_\mathrm{int}<1.0$ mag), but such systematic offsets mainly shift the intercept of the fitted relation rather than the slope, which is the quantity used to measure the CER. This approach, in which the color--color slope is used as a direct proxy for the CER without subtracting star-by-star intrinsic colors, has been widely adopted in earlier studies \citep[][and references therein]{2007ApJ...663.1069F, 2009ApJ...707...89G, 2009ApJ...707..510Z,2013ApJ...773...30W} and has also been applied in recent work on high-extinction isolated cloud cores \citep{li2024flattestinfraredextinctioncurve}. We therefore follow this methodology. The uncorrected fits, shown in the upper panels of Figure~\ref{fig:curvature}, exhibit tight but slightly curved correlations.

However, the upper panels of Figure~\ref{fig:curvature} show systematic curvature in the color--color relations when $Y-H>3.1$ mag. To characterize this high-extinction non-linearity quantitatively, we applied second-order (red) and third-order (green) polynomial fits. As discussed by \citet{Wang2019}, such curvature arises when fixed effective wavelengths are used in CER calculations. As stellar spectra are reddened by dense dust, their observed energy distributions shift toward longer wavelengths, reducing the filter-averaged extinction in each band. This effect is more pronounced for broader passbands, where the effective wavelength is more sensitive to the source spectrum and extinction. The evolution of filter-averaged extinction can be expressed as:

\begin{gather}
      A_{\lambda} = -2.5 \times \log\left( \frac{\int F_{\lambda}S_{\lambda}R_{\lambda}\,d\lambda}{\int F_{\lambda}S_{\lambda}\,d\lambda} \right)
      \label{eq1}
\end{gather}

where $F_\lambda$ is the intrinsic stellar flux, $S_\lambda$ is the filter transmission curve, and $R_\lambda$ is the wavelength-dependent attenuation factor set by the adopted extinction law and extinction normalization. According to this expression, a gradual reduction in filter-averaged extinction is unavoidable unless the filter bandwidth is infinitely narrow.

Furthermore, slight variations in effective wavelengths between different stellar types can lead to differences in derived relative extinctions, such as $A_\mathrm{VIS}/A_H$ and $A_J/A_H$ \citep{2023ApJ...956...26L}. In this work, we first calculate the unreddened effective wavelengths by:

\begin{gather}
      \lambda_{\text {eff, } 0}=\frac{\int \lambda F_\lambda S_\lambda d \lambda}{\int F_\lambda S_\lambda d \lambda }
      \label{eq2}
\end{gather}
Using PARSEC (PAdova and TRieste Stellar Evolution Code)\footnote[1]{https://stev.oapd.inaf.it/cgi-bin/cmd} isochrone models \citep{2012MNRAS.427..127B}, we adopt representative physical parameters for the red-giant population expected to dominate the selected reddened sequence: $\log g=2.0$, $\log T_{\rm eff}=3.64$, and [M/H]$=-0.1$. The resulting unreddened effective wavelengths are derived via Equation~\ref{eq2} as $\lambda_\mathrm{VIS}=0.714\,\mu$m, $\lambda_Y=1.072\,\mu$m, $\lambda_J=1.354\,\mu$m, and $\lambda_H=1.737\,\mu$m.

Finally, we convolve the corresponding model spectra with the filter transmission curves of each photometric system to obtain the correction terms, which are then applied to the color--color diagrams to remove the observed curvature. After this curvature correction, we repeat the linear regressions. As shown in the lower panels of Figure~\ref{fig:curvature}, the correction removes the curvature and produces significantly more linear color--color distributions.

\subsection{Relative extinction} \label{subsec:relative}
Using the CERs derived in the previous section, $k_\lambda = E(\lambda-H)/E(Y-H)$ ($\lambda={\rm VIS},J$), we convert them into relative extinctions $A_\lambda/A_H$ as
\begin{gather}
    A_\lambda / A_H=k_\lambda(A_Y / A_H-1)+1
    \label{eq3}
\end{gather}
This conversion requires an estimate of $A_Y / A_H$. Because NIR extinction is commonly described by a power law, $A_\lambda \propto \lambda^{-\alpha}$, the NIR CER can be expressed as a function of $\alpha$ by
\begin{gather}
    \frac{E(J-H)}{E(Y-H)}=\frac{\left(\frac{\lambda_H}{\lambda_J}\right)^\alpha-1}{\left(\frac{\lambda_H}{\lambda_Y}\right)^\alpha-1}
    \label{eq4}
\end{gather}
Using $E(J-H)/E(Y-H)=0.422 \pm 0.004$ from Section~\ref{subsec:CER} and the curvature-corrected effective wavelengths, Equation~\ref{eq4} yields a power-law index of $\alpha=1.57 \pm 0.06$, and thus $A_Y/A_H=2.13 \pm 0.18$ and $A_J/A_H=1.47\pm 0.11$. Finally, substituting $A_Y/A_H$ and $E({\rm VIS}-H)/E(Y-H)=2.855$ into Equation~\ref{eq3} gives $A_{\rm VIS}/A_H=4.23 \pm 0.24$. We therefore obtain the relative extinction law of LDN~1641 from the visible to the NIR.

\section{Discussion} \label{sec:discussion} 

To characterize extinction laws in different spatial environments, we select high- and low-extinction subsamples from the full sample in the final study region using $(Y-H)$ as an extinction proxy. This color approximately separates dense cores from diffuse outer regions. We define the two high-extinction core subregions with mean $(Y-H)>2.35$ mag as the high-extinction subsample, indicated by red dashed boundaries in the right panel of Figure~\ref{fig:herschel} and labeled 1--2. Similarly, we define the five low-extinction reference subregions with mean $(Y-H)<1.3$ mag as the low-extinction subsample, marked by black boundaries in the right panel of Figure~\ref{fig:herschel} and labeled 1--5. Because reddening signals are weaker in low-extinction conditions and intrinsic color dispersion has a relatively larger impact on fitting, we combine the five low-extinction reference subregions into a single low-extinction subsample for joint fitting to improve statistical precision and reduce biases caused by intrinsic color scatter.

\begin{figure*}[hbtp]
    \centering
    \includegraphics[width=1\linewidth]{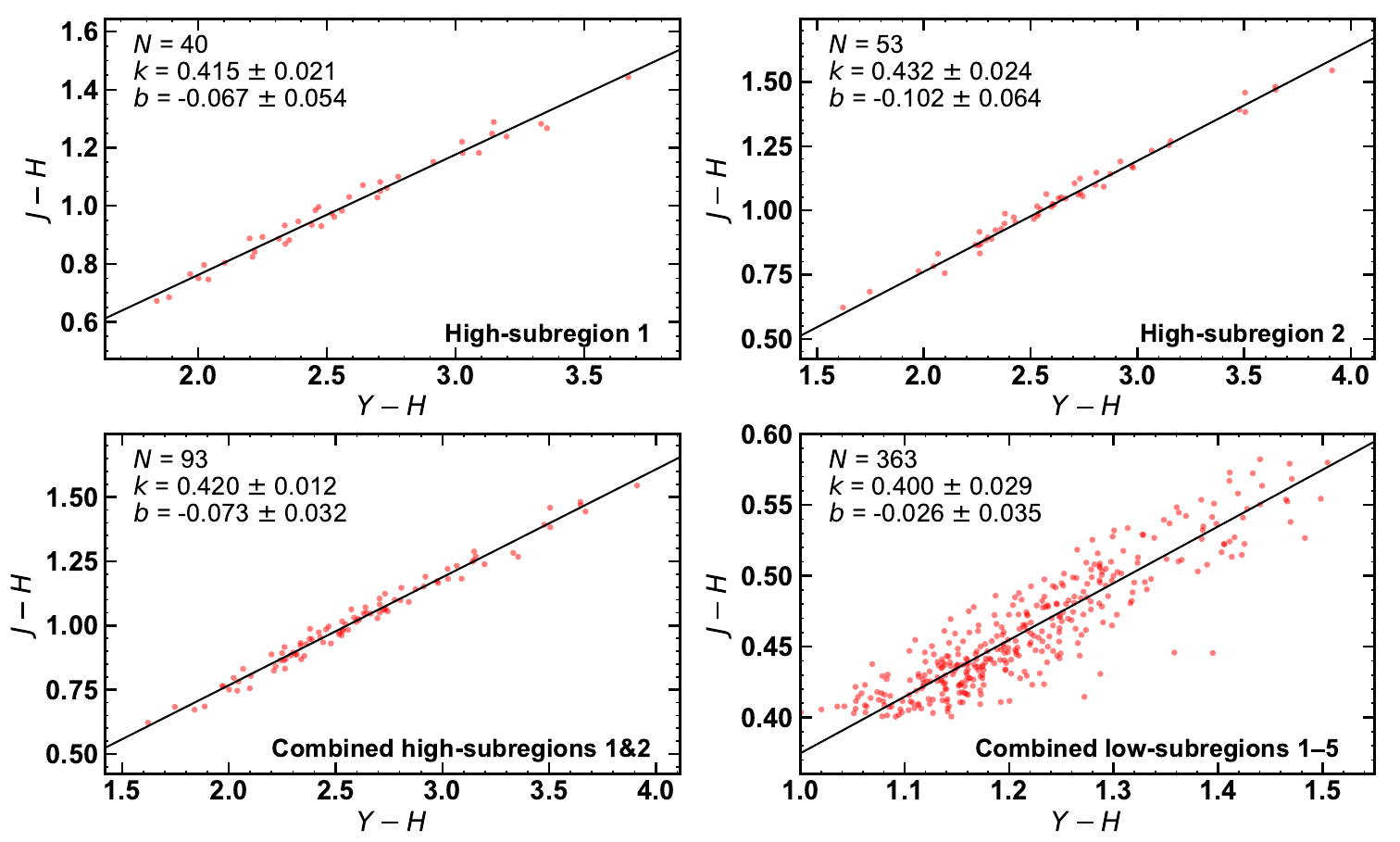}
    \caption{Color excess $E(J-H)$ versus $E(Y-H)$ diagrams for subregions with different extinction levels in LDN~1641. Top row: Individual high-extinction core subregions 1 (left) and 2 (right), selected with $(Y-H)>2.35$ mag. Bottom row: Combined high-extinction core subregions 1 and 2 (left) and the combined low-extinction reference subregions 1--5 (right, mean $(Y-H)<1.3$ mag). In each panel, red circles represent individual sources, and the black solid lines indicate the best-fitting linear relations; the number of sources and fit parameters are annotated in each panel.}
    \label{fig:special-region-fitting}
\end{figure*}
We apply the same method described in Section~\ref{sec:method} to the selected high- and low-extinction regions to derive CERs and the corresponding power-law index $\alpha$. Figure~\ref{fig:special-region-fitting} presents the $(J-H)$ versus $(Y-H)$ diagrams for high-extinction core subregions~1 and 2, the combined high-extinction core sample obtained by merging these two subregions, and the combined low-extinction reference sample obtained by merging reference subregions~1--5. All four samples exhibit clear linear relations. The combined high-extinction core sample provides a statistically more stable comparison with the combined low-extinction reference sample, and both the individual and combined high-extinction samples show systematically larger CERs than the low-extinction population. After converting the fitted $E(J-H)/E(Y-H)$ values into power-law indices, we obtain $\alpha=1.69 \pm 0.12$ and $1.41 \pm 0.14$ for high-extinction core subregions~1 and 2, respectively. The combined high-extinction core sample gives $\alpha=1.61 \pm 0.09$, whereas the combined low-extinction reference sample gives $\alpha=1.95 \pm 0.21$. The latter is consistent with the average NIR extinction index of the Milky Way \citep{Wang2019,2024ApJ...964L...3W}. The difference between the two combined samples is about 17\%, while the maximum difference between the individual subregions reaches about 27\%. These results suggest spatial variations in the NIR extinction curve within LDN~1641. In the high-extinction regions, the extinction curve is flatter, corresponding to smaller $\alpha$, which is consistent with expectations for enhanced large-grain populations in dense environments. 

To further test the environmental dependence indicated by the significantly reduced $\alpha$ in high-extinction regions, and to connect the Euclid results in dense cloud interiors with traditional NIR photometric studies, we introduce 2MASS $(J,H,K_S)$ data as an independent reference sample. Compared with 2MASS, Euclid provides greater photometric depth and higher spatial resolution in the $H$ band, allowing the detection of numerous faint sources along sightlines with $A_V\gtrsim 20$ mag and thereby effectively sampling high-extinction regions within the cloud. In contrast, the shallower depth of 2MASS means that most usable stars lie in the foreground or in low-extinction cloud edges, making it more suitable for characterizing average NIR extinction under diffuse or low-extinction conditions.

For this comparison, we select all 2MASS sources in the same sky area, yielding an initial sample of 2,344 sources. Following the quality-control and sample-selection criteria of \cite{Wang2019,2023ApJ...956...26L}, we require photometric uncertainties smaller than 0.03 mag in the $J$, $H$, and $K_S$ bands and restrict the magnitude range to 6--14 mag to avoid saturation and low signal-to-noise systematic effects; 328 sources remain after these cuts. Inspection of the color--magnitude diagram indicates contamination by asymptotic giant branch (AGB) stars. To reduce this contamination, we cross-match the 2MASS sample with the Spitzer Infrared Array Camera (IRAC) \citep{2004ApJS..154....1W,2004ApJS..154...10F} and Wide-field Infrared Survey Explorer (WISE) \citep{2010AJ....140.1868W} catalogs using a $1^{\prime\prime}$ radius. We remove likely AGB contaminants using mid-infrared color diagnostics, requiring the retained giant-star sequence to satisfy $[3.6]-[4.5] < 0.6$ mag, $[5.8]-[8.0] < 0.2$ mag \citep{2013ApJ...773...30W}, and $K_S-W3<1.5$; we also exclude sources that are clearly offset from the reddening direction in the $J-K_S$ versus $K_S$--WISE color--color planes. This procedure removes 76 AGB candidates and leaves 252 sources for the final 2MASS fit. The spatial distribution of these 2MASS sources shows that they lie mainly outside the selected high-extinction region defined from the Euclid $(Y-H)$ map, or along its lower-extinction edges (right panel of Figure~\ref{fig:herschel}), confirming that the 2MASS comparison traces low-extinction sightlines rather than the cloud interior.

The cleaned 2MASS sample follows a tight linear distribution in the $(H-K_S)$ versus $(J-H)$ diagram. A linear fit gives $E(H-K_S)/E(J-H)=0.467 \pm 0.010$, which corresponds to a NIR extinction power-law index of $\alpha=2.25 \pm 0.08$. This result is consistent, within the uncertainties, with $\alpha=1.95 \pm 0.21$ obtained for the combined low-extinction reference sample, but is substantially higher than $\alpha=1.61 \pm 0.09$ for the combined high-extinction core sample and $\alpha=1.41 \pm 0.14$ for Core~2, the densest high-extinction core subregion.
The 2MASS comparison therefore supports the conclusion that the NIR extinction curve in LDN~1641 becomes significantly flatter, with smaller $\alpha$, in dense high-extinction environments, consistent with substantial dust-grain growth under these conditions. A summary of the CER measurements and derived $\alpha$ values for all samples is provided in Table~\ref{tab:extinction_results}. As illustrated in Table~\ref{tab:extinction_results}, the derived power-law index $\alpha$ is sensitive to the CER $E(J-H)/E(Y-H)$. We attribute much of this uncertainty to systematic effects: sample-selection limitations may leave residual foreground or low-extinction contamination, and the statistical stability of the fits is limited by the sparse sample sizes in individual subregions. Despite the dispersion in $\alpha$ caused by these factors, the increase in the CER slope $k=E(J-H)/E(Y-H)$ from the cloud periphery to the core remains robust and supports grain growth in the densest regions.

\begin{table*}[ht]
\centering
\caption{Summary of CERs and $\alpha$ values}
\label{tab:extinction_results}
\begin{tabular}{lccc}
\hline
\hline
Sample/subregion & $E(J-H)/E(Y-H)$ & $E(\mathrm{VIS}-H)/E(Y-H)$ & $\alpha$ \\
\hline
Full sample & $0.422 \pm 0.004$ & $2.855 \pm 0.028$ & $1.57 \pm 0.06$ \\
High-extinction core subregion 1 & $0.415 \pm 0.021$ & $\cdots$ & $1.69 \pm 0.12$ \\
High-extinction core subregion 2 & $0.432 \pm 0.024$ & $\cdots$ & $1.41 \pm 0.14$ \\
Combined high-extinction core subregions 1 and 2 & $0.420 \pm 0.012$ & $\cdots$ & $1.61 \pm 0.09$ \\
Combined low-extinction reference subregions 1--5 & $0.400 \pm 0.029$ & $\cdots$ & $1.95 \pm 0.21$ \\
2MASS & $\cdots$ & $\cdots$ & $2.25 \pm 0.08$ \\
\hline
\end{tabular}
\tablefoot{CERs for Euclid bands are derived from curvature-corrected effective wavelengths. 
For the subregions, only the NIR CER $E(J-H)/E(Y-H)$ and the corresponding $\alpha$ are listed. 
For the 2MASS sample, the fitted CER is $E(H-K_S)/E(J-H)=0.467\pm0.010$, and the listed $\alpha$ value is obtained from this independent $(H-K_S)$ versus $(J-H)$ fit after AGB rejection.}
\end{table*}

\section{Summary} \label{sec:summary} 
Based on high-precision photometric data from Euclid Q1 VIS and NISP ($Y$, $J$, $H$), this work presents a systematic measurement of the optical-to-NIR extinction law in the Orion A dark cloud LDN~1641, with particular emphasis on its spatial variation in dense, high-extinction environments. We employ the color-excess method using $(\lambda-H)$ versus $(Y-H)$ color--color diagrams to derive CERs, and correct for possible effective-wavelength shifts at high extinction in order to obtain robust CERs and relative extinctions.

For the full sample in the final study region, we obtain a NIR extinction power-law index of $\alpha=1.57 \pm 0.06$, with relative extinctions $A_{\mathrm{VIS}}/A_H=4.23 \pm 0.24$, $A_Y/A_H=2.13 \pm 0.18$, and $A_J/A_H=1.47 \pm 0.11$. We further divide the region into high-extinction core subregions and low-extinction reference subregions according to $(Y-H)$, revealing a clear environmental dependence of the extinction law. The two high-extinction core subregions yield $\alpha$ values of $1.69 \pm 0.12$ and $1.41 \pm 0.14$, respectively, while the combined high-extinction sample gives $\alpha=1.61 \pm 0.09$ and the combined low-extinction reference sample gives $\alpha \simeq 1.95 \pm 0.21$, consistent with the average NIR extinction index of the Milky Way. We also use 2MASS $(J,H,K_S)$ data as an independent reference sample over the same sky area. After quality cuts and AGB rejection based on Spitzer and WISE mid-infrared colors, the final 2MASS sample contains 252 sources and is mainly located outside the selected high-extinction region. It gives $E(H-K_S)/E(J-H)=0.467\pm0.010$, corresponding to $\alpha=2.25 \pm 0.08$, and therefore provides a low-extinction NIR reference. Taken together, these results show that the NIR extinction curve in LDN~1641 becomes significantly flatter with increasing environmental density: in dense high-extinction regions, $\alpha$ systematically decreases, with a difference of about 17\% between the combined high- and low-extinction samples and up to 27\% between individual subregions. This result indicates clear spatial evolution of the extinction law within the molecular cloud. With the broader sky coverage provided by future Euclid data releases, it will be possible to characterize the extinction law in high-extinction regions of the Galactic plane over larger areas and to greater depth, thereby refining our understanding of dust evolution.

\begin{acknowledgements}
We thank the anonymous referee for constructive comments and suggestions that helped improve the manuscript.
This work is supported by the National Natural Science Foundation of China (NSFC) through the projects 12373028, 12322306, 12373035, 12133002, and 12173047. S. W. and X.C. acknowledge support from the Youth Innovation
Promotion Association of the Chinese Academy of Sciences (CAS) with Nos. 2023065 and 2023055. This work has made use of the Euclid Q1 data from the European Space Agency (ESA) mission Euclid, available at https://www.cosmos.esa.int/en/web/euclid/q1-data.
\end{acknowledgements}

\bibliographystyle{aa}
\bibliography{bibliography}

\end{document}